\documentclass[12pt]{article}

\usepackage{amsmath,amssymb}
\usepackage{graphicx}% Include figure files
\usepackage{caption}
\usepackage{color}% Include colors for document elements
\usepackage{dcolumn}% Align table columns on decimal point
\usepackage{bm}% bold math
\usepackage[numbers,super,comma,sort&compress]{natbib}
\usepackage{float}
\usepackage{hyperref} % hyperref must be loaded before apacite
\hypersetup{colorlinks=true,citecolor=black,urlcolor=blue}%\usepackage[nolists, nomarkers, figuresfirst]{endfloat}

\definecolor{background-color}{gray}{0.98}

\def\rz{\ifmmode{I\hskip -3pt R}
    \else{\hbox{$I\hskip -3pt R$}}\fi} %reelle Zahlen

\def\bb{{\boldsymbol{b}}}
\def\bk{{\boldsymbol{k}}}

\def\bt{{\boldsymbol{t}}}

\def\bx{{\boldsymbol{x}}}

\def\bz{{\boldsymbol{z}}}
\def\bA{{\boldsymbol{A}}}
\def\bE{{\boldsymbol{E}}}
\def\bL{{\boldsymbol{L}}}

\def\bR{{\boldsymbol{R}}}
\def\bS{{\boldsymbol{S}}}
\def\bT{{\boldsymbol{T}}}
\def\bV{{\boldsymbol{V}}}
\def\bX{{\boldsymbol{X}}}
\def\bY{{\boldsymbol{Y}}}
\def\bZ{{\boldsymbol{Z}}}
\def\bmu{\boldsymbol{\mu}}
\def\hbmu{\hat{\boldsymbol{\mu}}}
\def\bSigma{\boldsymbol{\Sigma}}
\def\hbSigma{\hat{\boldsymbol{\Sigma}}}
\DeclareMathOperator*{\argmin}{argmin}
\def\bLambda{\boldsymbol{\Lambda}}

\title{Minimum Covariance Determinant and Extensions}
\author{Mia Hubert\thanks{Department of 
  Mathematics, KU Leuven, Celestijnenlaan 
	200B, BE-3001 Leuven, Belgium}, 
	Michiel Debruyne\thanks{Dexia Bank, 
	Belgium}, 
	Peter J. Rousseeuw\footnotemark[1]}
\date{\today}
\begin{document}
\maketitle

\thispagestyle{empty}

\begin{center}
%\subsubsection*{\small Article Type:}
%Advanced Review
%%The Article Type denotes the intended level of readership for your article. An Editor may have mentioned a specific Article Type in your invitation letter; if so, please let them know if you think a different Article Type better suits your topic.

\hfill \break
\thanks

\subsubsection*{Abstract}
\begin{flushleft}
The Minimum Covariance Determinant (MCD) method 
is a highly robust estimator of multivariate 
location and scatter, for which a fast algorithm 
is available. Since estimating the covariance 
matrix is the cornerstone of many multivariate 
statistical methods, the MCD is an important 
building block when developing robust multivariate 
techniques. It also serves as a convenient and 
efficient tool for outlier detection.   

The MCD estimator is reviewed, along with its 
main properties such as affine equivariance, 
breakdown value, and influence function. 
We discuss its computation, and list applications 
and extensions of the MCD in applied and
methodological multivariate statistics. Two 
recent extensions of the MCD are described. The 
first one is a fast deterministic algorithm which 
inherits the robustness of the MCD while being 
almost affine equivariant. The second is tailored 
to high-dimensional data, possibly with more 
dimensions than cases, and incorporates 
regularization to prevent singular matrices.
\end{flushleft}
\end{center}

% makes references listed with 1., 2., etc:  
\makeatletter
\renewcommand\@biblabel[1]{#1.}
\makeatother
\bibliographystyle{unsrtnat} % This is WIREs `Vancouver style'
\renewcommand{\baselinestretch}{1.5}
\normalsize
\clearpage

\newpage
\section*{\sffamily \Large INTRODUCTION} 
The Minimum Covariance Determinant (MCD) estimator 
is one of the first affine equivariant and highly 
robust estimators of multivariate location and 
scatter \cite{Rousseeuw:LMS,Rousseeuw:MVE}. 
Being resistant to outlying observations makes the 
MCD very useful for outlier detection. Although 
already introduced in 1984, its main use has only 
started since the construction of the 
computationally efficient FastMCD algorithm of 
\cite{Rousseeuw:FastMCD}
in 1999. % remove this year if `full' citations are used.
Since then, the MCD has been applied in numerous 
fields such as medicine, finance, image analysis
and chemistry. Moreover the MCD has also been used 
to develop many robust multivariate techniques, 
among which robust principal component analysis, 
factor analysis and multiple regression. 
Recent modifications of the MCD include a 
deterministic algorithm and a regularized version 
for high-dimensional data. 

\section*{\sffamily \Large DESCRIPTION OF THE MCD ESTIMATOR} 
\subsection*{\sffamily \large Motivation}
In the multivariate location and scatter setting 
the data are stored in an $n \times p$ data matrix
$\bX=(\bx_1,\ldots,\bx_n)'$ with 
$\bx_i=(x_{i1},\ldots,x_{ip})'$ the $i$-th 
observation, so $n$ stands for the
number of objects and $p$ for the number of variables. 
We assume that the observations are sampled from an
elliptically symmetric unimodal distribution
with unknown parameters $\bmu$ and $\bSigma$, where
$\bmu$ is a vector with $p$ components and $\bSigma$
is a positive definite $p \times p$ matrix.
To be precise, a multivariate distribution is called
elliptically symmetric and unimodal if there exists a 
strictly decreasing real function $g$ such that the 
density can be written in the form
\begin{align}\label{eq:ell}
  f(\bx)=\frac{1}{\sqrt{|\bSigma|}}\;
	g(d^2(\bx,\bmu,\bSigma))
\end{align}
in which the \emph{statistical distance} 
$d(\bx,\bmu,\bSigma)$ is given by
\begin{align}\label{eq:d}
   d(\bx,\bmu,\bSigma)=
	\sqrt{(\bx-\bmu)'\bSigma^{-1}(\bx-\bmu)}\ \ .
\end{align} 

\begin{figure}[H]
\begin{center}
\includegraphics[keepaspectratio=true,scale=0.9]
                {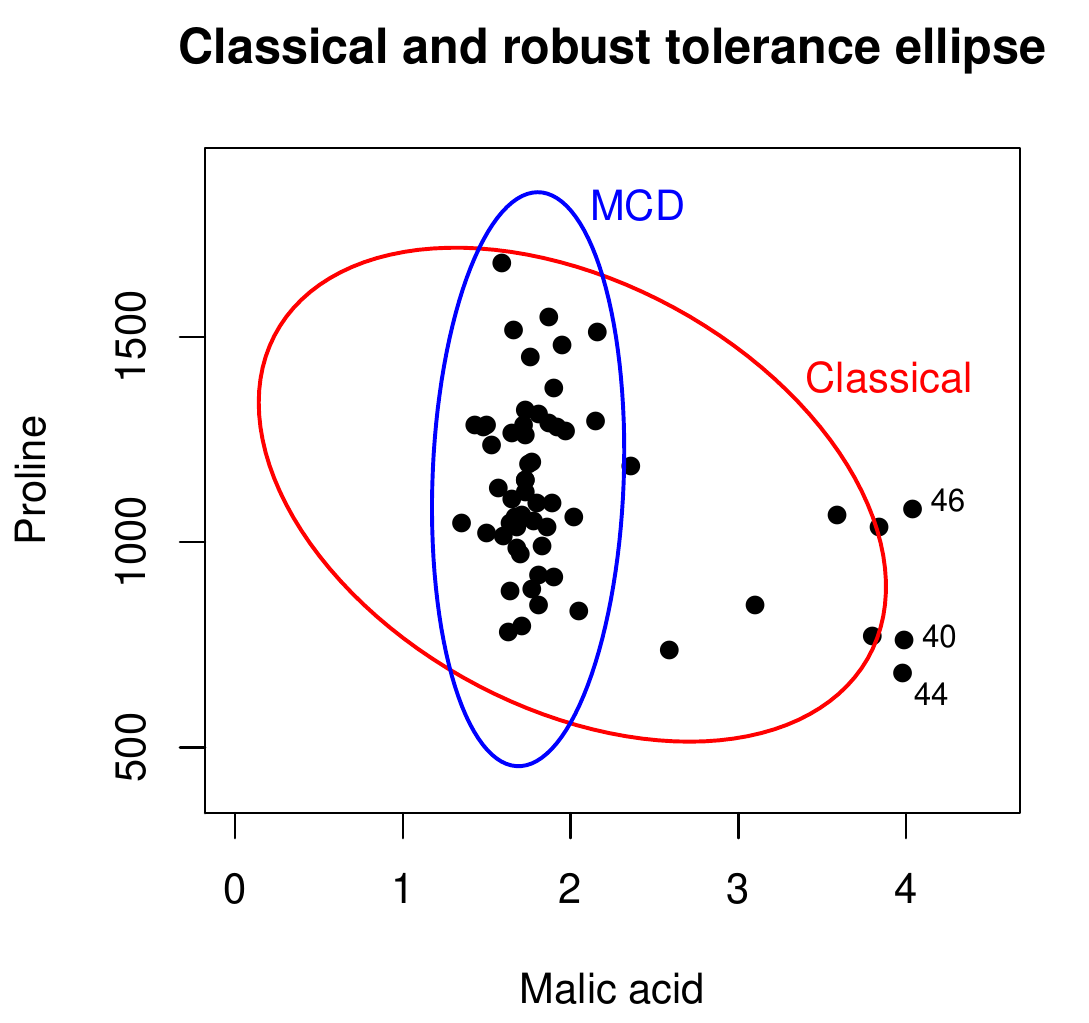}	
\caption{Bivariate wine data: tolerance ellipse
         of the classical mean and covariance
				 matrix (red), and that of the robust 
				 location and scatter matrix (blue).}
\label{fig:tolellipswine} 
\end{center}
\end{figure}

To illustrate the MCD, we first consider the 
wine data set available
in \cite{Hettich:UCIKDDArchive} and also analyzed 
in \cite{Maronna:RobStat}. 
This data set contains the quantities of 13 
constituents found in three types of Italian wines. 
We consider the first group containing 59 wines, 
and focus on the constituents `Malic acid' and 
`Proline'. 
This yields a bivariate data set, i.e. $p=2$.
A scatter plot of the data is shown in
Figure \ref{fig:tolellipswine}, in which we see 
that the points on the lower right hand side of 
the plot are outlying relative to the majority of 
the data.

\begin{figure}[H]
\begin{center}
\vskip0.1in
\begin{tabular}{cc}
\includegraphics[keepaspectratio=true,scale=0.725]
                {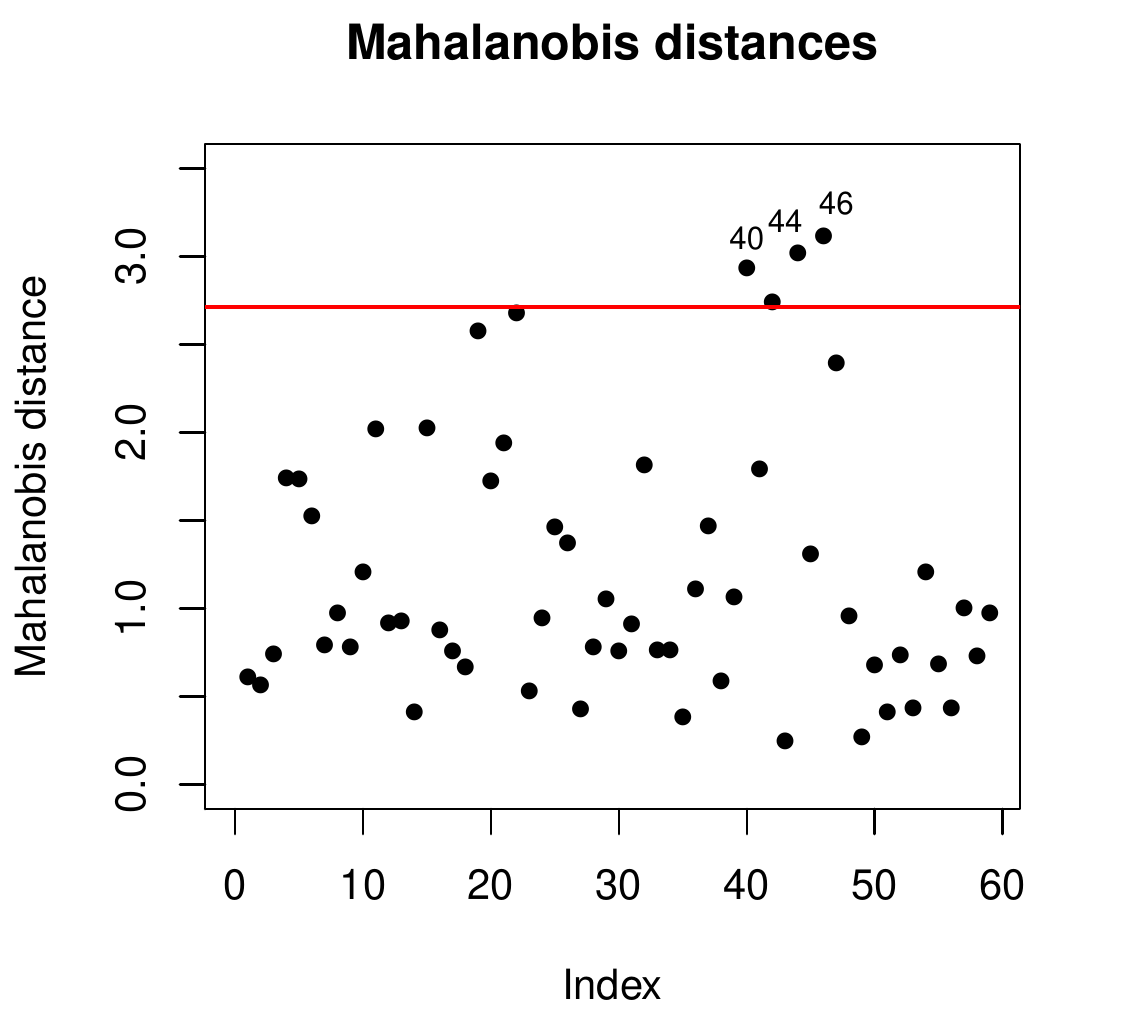} &
\includegraphics[keepaspectratio=true,scale=0.725]
                {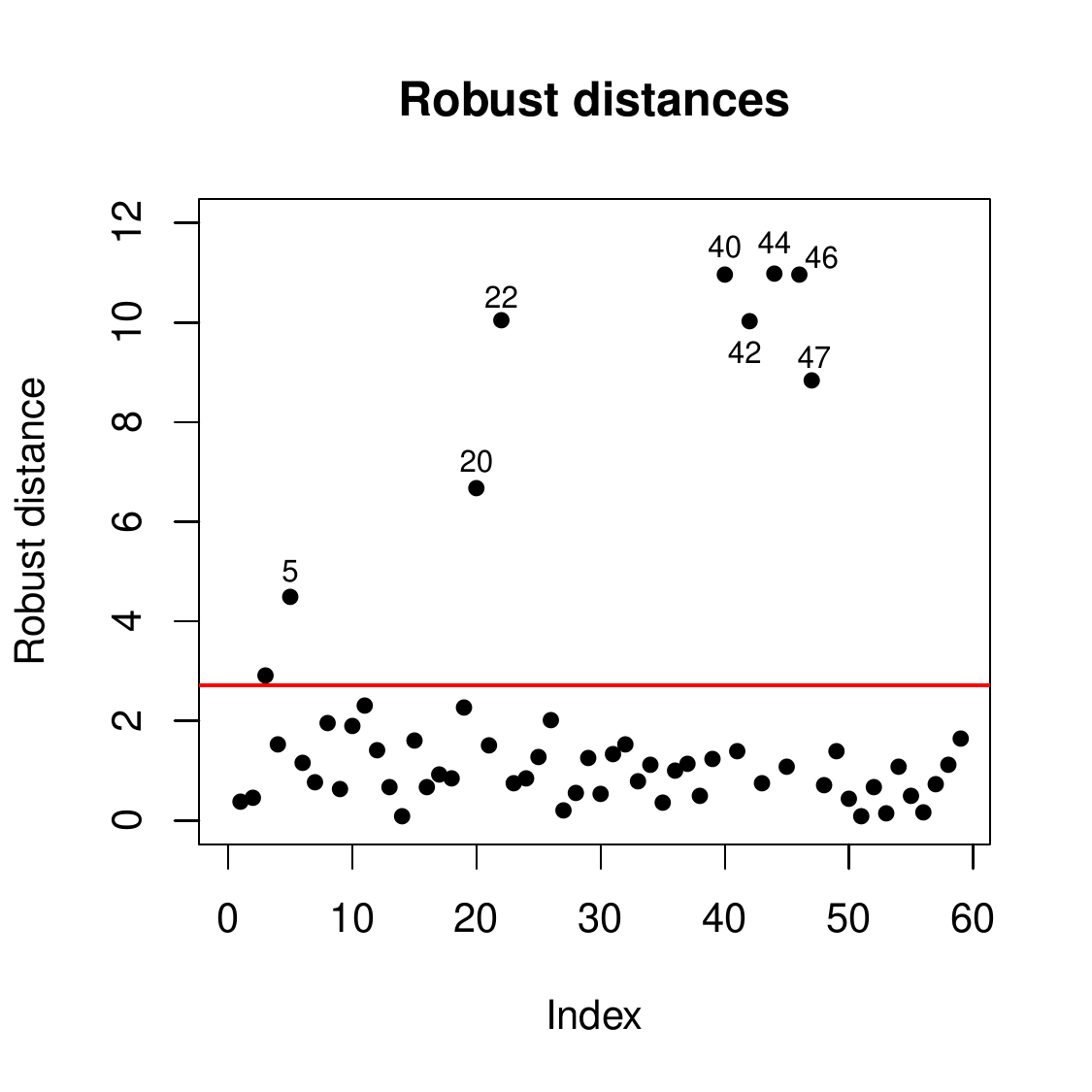} \\
(a) & (b)	\\										
\end{tabular}
\caption{(a) Mahalanobis distances and (b) 
         robust distances for the bivariate
				 wine data.} 
\label{fig:wined}
\end{center}
\end{figure}

In the figure we see two ellipses.
The classical tolerance ellipse is defined as the 
set of $p$-dimensional points $\bx$ whose 
{\it Mahalanobis distance}
\begin{equation}
  \text{MD}(\bx) =
	d(\bx,\bar{\bx},\mbox{Cov}(X)) = 
	\sqrt{(\bx -\bar{\bx})' \mbox{Cov}(X)^{-1}
	(\bx -\bar{\bx})}  
\label{eq:md}
\end{equation}
equals $\sqrt{\chi^2_{p,0.975}}$. 
Here $\bar{\bx}$ is the sample mean and 
$\mbox{Cov}(X)$ the sample covariance 
matrix. The Mahalanobis distance 
$\text{MD}(\boldsymbol{x}_i)$ should tell us how 
far away $\boldsymbol{x}_i$ is from the center of 
the data cloud, relative to its size and shape.  
In Figure \ref{fig:tolellipswine} we see that 
the red tolerance ellipse tries to encompass 
all observations. 
Therefore none of the Mahalanobis distances 
is exceptionally large, as we can see in 
Figure \ref{fig:wined}(a).
Based on Figure \ref{fig:wined}(a) alone we 
would say there are only three mild outliers
in the data (we ignore borderline cases).
 
On the other hand, the robust tolerance ellipse
is based on the robust distances
\begin{equation}
  \text{RD}(\bx) = d(\bx, \hbmu_{MCD}, 
	\hat{\bSigma}_{MCD})
\label{eq:rd}
\end{equation}
where $\hbmu_{MCD}$ is the MCD estimate of 
location and $\hat{\bSigma}_{MCD}$ is the MCD 
covariance estimate, which we will explain soon.
In Figure \ref{fig:tolellipswine} we see that the
robust ellipse (in blue) is much smaller and only 
encloses the regular data points. 
The robust distances shown in 
Figure \ref{fig:wined}(b) now clearly expose 8 
outliers.

This illustrates the \textit{masking effect}: 
the classical estimates can be so strongly 
affected by contamination that diagnostic tools 
such as the Mahalanobis distances become unable
to detect the outliers. To avoid masking we 
instead need reliable estimators that can resist 
outliers when they occur. The MCD is such a 
robust estimator.

\subsection*{\sffamily \large Definition}  
The raw Minimum Covariance Determinant (MCD) 
estimator with tuning constant
$n/2\leqslant h\leqslant n$ is
$(\hat{\bmu}_0,\hat{\bSigma}_0)$ where
\begin{enumerate}
\item the location estimate $\hat{\bmu}_0$ is 
 the mean of the $h$ observations for which the
 determinant of the sample covariance matrix is 
 as small as possible;
\item the scatter matrix estimate 
 $\hat{\bSigma}_0$ is the corresponding 
 covariance matrix multiplied by a consistency 
 factor $c_0$.
\end{enumerate}

Note that the MCD estimator can only be computed 
when $h > p$, otherwise the covariance matrix of 
any $h$-subset has determinant zero, so we need
at least $n > 2p$. To avoid excessive noise
it is however recommended that $n > 5p$, so that 
we have at least 5 observations per dimension.
(When this condition is not satisfied one can
instead use the MRCD method \eqref{eq:MRCD} 
described near the end of this article.) 
To obtain consistency at the normal distribution,
the consistency factor $c_0$ equals 
$\alpha/F_{\chi^2_{p+2}}(q_\alpha)$ 
with $\alpha=\lim_{n \to \infty} h(n)/n$,
and $q_\alpha$ the $\alpha$-quantile of 
the $\chi^2_p$ distribution \cite{Croux:IFMCD}. 
Also a finite-sample correction factor can be 
incorporated \cite{Pison:Corfac}.

Consistency of the raw MCD estimator of location and scatter at elliptical models, as well as asymptotic normality
of the MCD location estimator has been proved in \cite{Butler:AsymptMCD}. Consistency and asymptotic normality of the 
MCD covariance matrix at a broader class of distributions is derived in \cite{Cator:AsympMCD,Cator:Infl}. 

The MCD estimator is the most robust when 
taking $h=[(n+p+1)/2]$ where $[a]$ is the 
largest integer $\leqslant a$.
At the population level this corresponds to
$\alpha=0.5$. But unfortunately the MCD then 
suffers from low efficiency at the normal model. 
For example, if $\alpha=0.5$ the asymptotic 
relative efficiency of the diagonal elements of 
the MCD scatter matrix relative to the
sample covariance matrix is only 6\% when $p=2$, 
and 20.5\% when $p=10$. 
This efficiency can be increased by considering 
a higher $\alpha$ such as $\alpha=0.75$. This 
yields relative efficiencies of 26.2\% for $p=2$ 
and 45.9\% for $p=10$ (see \cite{Croux:IFMCD}). 
On the other hand this choice of $\alpha$ 
diminishes the robustness to possible outliers.

In order to increase the efficiency while retaining 
high robustness one can apply a weighting step
\cite{Lopuhaa:BDP,Lopuhaa:RewEst}. For the MCD 
this yields the estimates
\begin{align}
\label{eq:rewMCD}	
\begin{split} % to assign 1 label to 2 formulas
  \hat{\bmu}_{MCD} & = \frac{\sum_{i=1}^nW(d_i^2)
	     \bx_i}{\sum_{i=1}^nW(d^2_i)}\\
  \hat{\bSigma}_{MCD} & = c_1 \frac{1}{n}
	   \sum_{i=1}^nW(d_i^2)(\bx_i-\hat{\bmu}_{MCD})
		(\bx_i-\hat{\bmu}_{MCD})'
\end{split}
\end{align}
with $d_i=d(\bx,\hbmu_0,\hat{\bSigma}_0)$ 
and $W$ an appropriate weight function.
The constant $c_1$ is again a consistency 
factor. A simple yet effective choice for $W$ 
is to set it to 1 when the robust distance is
below the cutoff $\sqrt{\chi^2_{p,0.975}}$
and to zero otherwise, that is,
$W(d^2)=I(d^2 \leqslant \chi^2_{p,0.975})$. 
This is the default choice in the
current implementations in R, SAS, Matlab and S-PLUS. If we
take $\alpha=0.5$ this weighting step increases the efficiency to 45.5\% for $p=2$ and to 82\% for $p=10$. In the
example of the wine data (Figure \ref{fig:tolellipswine}) we applied the weighted MCD estimator with $\alpha=0.75$, but the results were similar
for smaller values of $\alpha$.

Note that one can construct a robust correlation
matrix from the MCD scatter matrix. 
The robust correlation between variables $X_i$ 
and $X_j$ is given by
\begin{equation*}
r_{ij} = \frac{s_{ij}}{\sqrt{s_{ii} s_{jj}}}
\end{equation*}
with $s_{ij}$ the $(i,j)$-th element of the MCD 
scatter matrix.
In Figure \ref{fig:tolellipswine} the MCD-based
robust correlation is $0.10 \approx 0$ because 
the majority
of the data do not show a trend, whereas the 
classical correlation of $-0.37$ was caused by
the outliers in the lower right part of the plot.

\subsection*{\sffamily \large Outlier detection}
As already illustrated in Figure \ref{fig:wined}, the robust MCD estimator is very useful to detect outliers in
multivariate data. As the robust distances \eqref{eq:rd} are not sensitive to the masking effect, they can be used
to flag the outliers
\cite{Rousseeuw:Diagnostic,Cerioli:outliers}. 
This is crucial for data sets in more than three dimensions, which are difficult to visualize.

\begin{figure}[H]
\begin{center}
\includegraphics[keepaspectratio=true,scale=0.9]
                {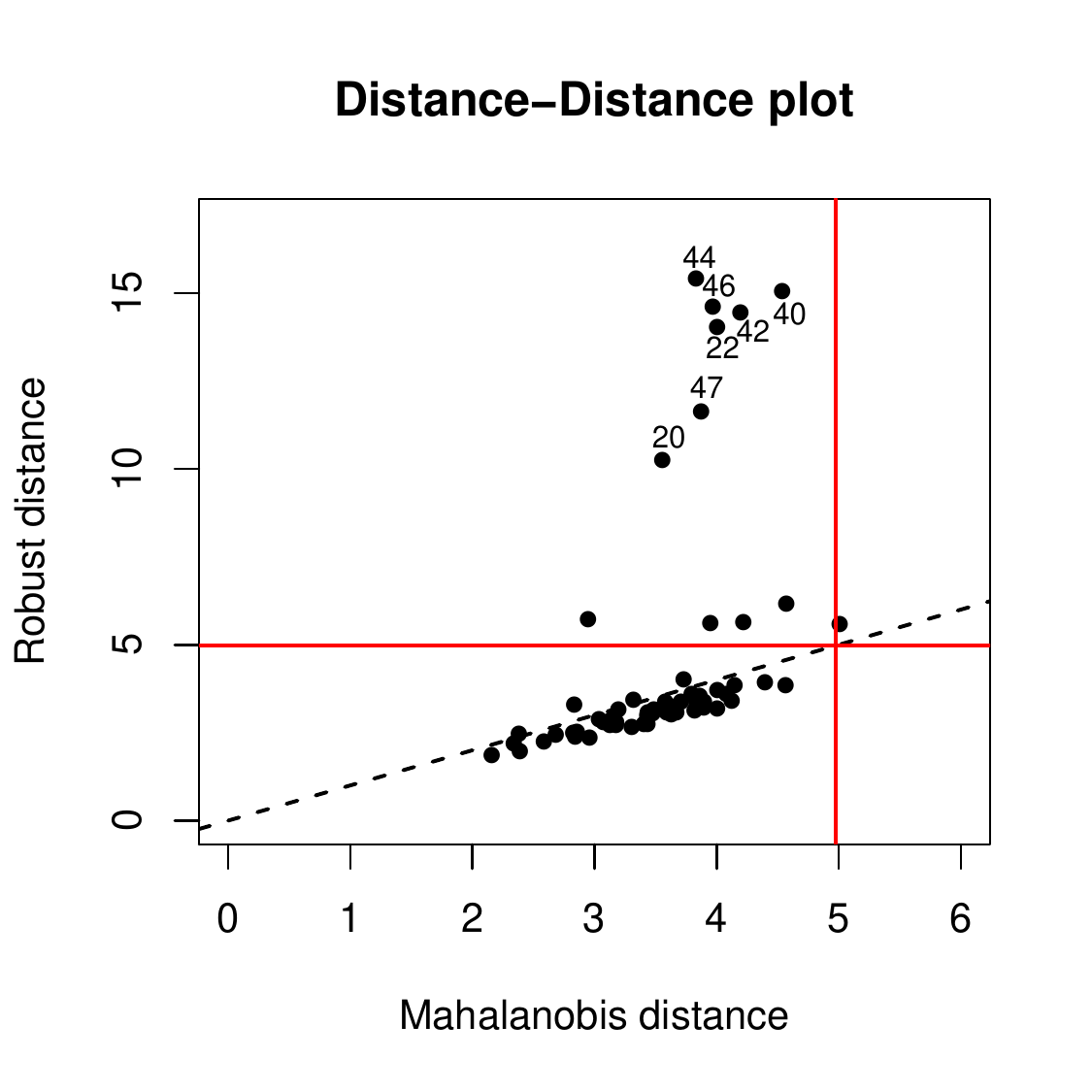}	
\caption{Distance-distance plot of the
         full 13-dimensional wine data set.} 
\label{fig:winedd}								
\end{center}
\end{figure}

We illustrate the outlier detection potential 
of the MCD on the full wine data set, with 
all $p=13$ variables. The
\textit{distance-distance plot} of 
\cite{Rousseeuw:FastMCD} in Figure 
\ref{fig:winedd} shows the robust distances 
based on the MCD versus the classical 
distances \eqref{eq:md}. From the robust 
analysis we see that seven observations 
clearly stand out (plus some mild outliers), 
whereas the classical analysis does not flag 
any of them.

Note that the cutoff value $\sqrt{\chi^2_{p,0.975}}$ is based on the asymptotic distribution of the robust
distances, and often flags too many observations as outlying. For relatively small $n$ the true distribution of the robust distances can be
better approximated by an $F$-distribution, see \cite{Hardin:RD}.

\section*{\sffamily \Large PROPERTIES} 

\subsection*{\sffamily \large Affine equivariance}

The MCD estimator of location and scatter 
is \emph{affine equivariant}. This means 
that for any nonsingular $p \times p$ 
matrix $\bA$ and any $p$-dimensional column 
vector $\bb$ it holds that
\begin{align}
\label{eq:affine}
 \hat{\bmu}_{MCD}(\bX \bA'+\mathbf{1}_n \bb')& =\hat{\bmu}_{MCD}(\bX)\bA'+\bb \\
 \hat{\bSigma}_{MCD}(\bX \bA'+\mathbf{1}_n \bb')& =\bA\hat{\bSigma}_{MCD}(\bX)\bA'
\end{align}
where the vector $\mathbf{1}_n$ is 
$(1,1,\dots,1)'$ with $n$ elements. 
This property follows from the fact that 
for each subset $H$ of $\{1,2,\ldots,n\}$
of size $h\,$ and corresponding data set 
$\bX_H\,$, the determinant of the
covariance matrix of the transformed data 
equals
$$|\bS(\bX_H \bA')|= |\bA \bS(\bX_H) \bA'|
   = |\bA|^2 |\bS(\bX_H)|.$$
Therefore, transforming an $h$-subset with
lowest determinant yields an $h$-subset
$\bX_H \bA'$ with lowest determinant among
all $h$-subsets of the transformed data set
$\bX \bA'\,$, and its covariance matrix is  
transformed appropriately. The affine
equivariance of the raw MCD location 
estimator follows from the equivariance of 
the sample mean. Finally we note that the 
robust distances 
$d_i=d(\bx,\hbmu_0,\hat{\bSigma}_{0})$ are 
affine \emph{invariant}, meaning they stay 
the same after transforming the data, which 
implies that the weighted estimator is 
affine equivariant too.

Affine equivariance implies that the estimator transforms well under any non-singular reparametrization of the
space in which the $\bx_i$ live. Consequently, the data might be rotated, translated or rescaled (for example through a
change of measurement units) without affecting the outlier detection diagnostics.

The MCD is one of the first high-breakdown 
affine equivariant estimators of location 
and scatter, and was only preceded by the 
Stahel-Donoho 
estimator \cite{Stahel:SDest,Donoho:Depth}. 
Together with the MCD also the Minimum
Volume Ellipsoid estimator was introduced 
\cite{Rousseeuw:LMS,Rousseeuw:MVE} which is
equally robust but not asymptotically normal,
and is harder to compute than the MCD.

\subsection*{\sffamily \large Breakdown value} 
The breakdown value of an estimator is the 
smallest fraction of observations that need 
to be replaced (by arbitrary values) to make
the estimate useless. 
For a multivariate {\it location} estimator
$T_n$ the breakdown value is defined as
\begin{align*}\label{eq:fsbp}
  \varepsilon^*_n(T_n;\bX_n)=\frac{1}{n}
	\min\left\{m:
	\sup ||T_n(\bX_{n,m}) - T_n(\bX_n)|| =
  +\infty\right\}
\end{align*}
where $1 \leqslant m \leqslant n$ and
the supremum is over all data sets
$\bX_{n,m}$ obtained by replacing any $m$ 
data points 
$\bx_{i_1},\hdots,\bx_{i_m}$ of $\bX_n$ by 
arbitrary points.

For a multivariate {\it scatter} 
estimator $C_n$ we set
\begin{align*}
  \varepsilon^*_n(C_n;\bX_n)=\frac{1}{n}
	\min\{ m:\sup\; 
	\max_i|\log(\lambda_i(C_n(\bX_{n,m})))-
	\log(\lambda_i(C_n(\bX_n)))|
	 = +\infty \}
\end{align*} 
with $\lambda_1(C_n) \geqslant \hdots
 \geqslant \lambda_p(C_n) > 0$ the 
eigenvalues of $C_n\,$.
This means that we consider a scatter estimator 
to be broken when $\lambda_1$ can become 
arbitrarily large (`explosion') and/or 
$\lambda_p$ 
can become arbitrary close to $0$ (`implosion').
Implosion is a problem because it makes the
scatter matrix singular whereas in many 
situations its inverse is required, e.g. 
in \eqref{eq:rd}.

Let $k(\bX_n)$ denote the highest number of 
observations in the data set that lie on an 
affine hyperplane in $p$-dimensional space,
and assume $k(\bX_n) < h$. Then the raw MCD 
estimator of location and scatter
satisfies \cite{Roelant:MWCD}
\begin{equation}
  \varepsilon^*_n(\hat{\bmu}_0;\bX_n)=
	\varepsilon^*_n(\hat{\bSigma}_0;\bX_n)= 
	\frac{\min(n-h+1,h-k(\bX_n))}{n}\;\;.
\end{equation}

If the data are sampled from a continuous 
distribution, then almost surely 
$k(\bX_n) = p$ which is called
\textsl{general position}. Then
$\varepsilon^*_n(\hat{\bmu}_0;\bX_n)=
\varepsilon^*_n(\hat{\bSigma}_0;\bX_n)= 
\min(n - h + 1, h - p)/n$, and
consequently any $[(n + p)/2] \leqslant h 
\leqslant [(n + p + 1)/2]$ gives the 
breakdown value $[(n - p + 1)/2]$. This is 
the highest possible breakdown value 
for affine equivariant scatter
estimators \cite{Davies:AsymptSest} at data 
sets in general position. Also for affine 
equivariant location estimators the upper 
bound on the breakdown value
is $[(n - p + 1)/2]$ under natural regularity 
conditions \cite{Rousseeuw:Discussion}. 
Note that in the limit
$\lim_{n \to \infty} \varepsilon^*_n = 
\min(1-\alpha,\alpha)$ which is maximal 
for $\alpha=0.5$.

Finally we note that the breakdown value 
of the weighted MCD estimators 
$\hat{\bmu}_{MCD}$ and $\hat{\bSigma}_{MCD}$
is not lower than the breakdown value of 
the raw MCD estimator, as long as the weight 
function $W$ used in \eqref{eq:rewMCD} is 
bounded and becomes zero for large 
$d_i\,$, see \cite{Lopuhaa:BDP}.

\subsection*{\sffamily \large Influence function}
The influence 
function \cite{Hampel:IFapproach} of an 
estimator measures the effect of a small
(infinitesimal) fraction of outliers placed
at a given point.
It is defined at the population level hence 
it requires the functional form of the 
estimator $T$, which maps a distribution $F$ 
to a value $T(F)$ in the parameter space. 
For multivariate location this parameter 
space is $\rz^p$, whereas for multivariate 
scatter the parameter space is
the set of all positive semidefinite 
$p \times p$ matrices. The influence function 
of the estimator $T$ at the distribution $F$ 
in a point $\bx$ is then defined as
\begin{equation}
  IF(\bx,T,F) = \lim_{\varepsilon \to 0} 
	\frac{T(F_\varepsilon)-T(F)}{\varepsilon} 
	\label{eq:IF}
\end{equation}
with $F_\varepsilon=(1-\varepsilon)F+
 \varepsilon \Delta_x$ a contaminated 
distribution with point mass in $\bx$.

The influence function of the raw and the 
weighted MCD has been computed in 
\cite{Croux:IFMCD,Cator:Infl} and turns out
to be bounded. This is a desirable property 
for robust estimators, as it limits the 
effect of a small fraction of outliers on
the estimate.
At the standard multivariate normal 
distribution, the influence function of the 
MCD location estimator becomes zero for all 
$\bx$ with $\|\bx\|^2 > \chi^2_{p,\alpha}$ 
hence far outliers do not influence the
estimates at all. The same happens with the 
off-diagonal elements of the MCD scatter 
estimator. On the other hand, the influence 
function of the diagonal elements remains 
constant (different from zero) when 
$\|\bx\|^2$ is sufficiently large. Therefore 
the outliers still have a bounded influence 
on the estimator. All these influence
functions are smooth, except at those $\bx$ 
with $\|\bx\|^2 = \chi^2_{p,\alpha}\,$. 
The weighted MCD estimator has an 
additional jump in 
$\|\bx\|^2 = \chi^2_{p,0.975}$ due to the 
discontinuity of the weight function, but
one could use a smooth weight function
instead.

\subsection*{\sffamily \large Univariate MCD}
For univariate data $x_1,\ldots,x_n$
the MCD estimates reduce 
to the mean and the standard deviation of the 
$h$-subset with smallest variance. They can 
be computed in O($n \log n$) time by sorting
the observations and only considering 
contiguous $h$-subsets so that their means 
and variances can be calculated
recursively \cite{Rousseeuw:RobReg}. 
Their consistency and asymptotic normality is 
proved in \cite{Butler:uniMCD,Rousseeuw:MVE}. 
For $h=[n/2]+1$ the MCD location estimator 
has breakdown value $[(n+1)/2]/n$ and the 
MCD scale estimator has $[n/2]/n$. 
These are the highest breakdown values that 
can be attained by univariate affine 
equivariant estimators \cite{Croux:HBscale}. 
The univariate MCD estimators also have 
bounded influence functions,
see \cite{Croux:IFMCD} for details. 
Their maximal asymptotic bias is studied 
in \cite{Croux:BiasScale,Croux:BiasLoc} 
as a function of the contamination fraction.

Note that in the univariate case the 
MCD estimator corresponds to the Least 
Trimmed Squares (LTS) regression 
estimator \cite{Rousseeuw:LMS}, which is
defined by
\begin{equation}
  \hat{\beta}_{LTS} = \argmin_\mu 
  \sum_{i=1}^h (r^2_\beta)_{i:n}
  \label{eq:LTS}
\end{equation}
where $(r^2_\beta)_{1:n} \leqslant
 (r^2_\beta)_{2:n} \leqslant \ldots 
 \leqslant (r^2_\beta)_{n:n}$ are the 
ordered squared residuals. For
univariate data these residuals are
simply $(r_{\beta})_i = x_i-\beta\,$.

\section*{\sffamily \Large COMPUTATION} 
The exact MCD estimator is very hard to 
compute, as it requires the evaluation of 
all $\binom{n}{h}$ subsets of size $h$. 
Therefore one switches to an approximate 
algorithm such as the FastMCD algorithm 
of \cite{Rousseeuw:FastMCD} which is quite
efficient. The key
component of the algorithm is the C-step:
\vskip0.2in
\noindent {\bf Theorem.}
 {\it
   Take $\bX=\{\boldsymbol{x}_1,\dots,\boldsymbol{x}_n\}$ and let
   $H_1 \subset \{1,\dots,n\}$ be a subset of size $h$. Put
   $\boldsymbol{\hat{\mu}}_1$ and
   $\boldsymbol{\hat{\Sigma}}_1$ the empirical mean and covariance matrix of the data in $H_1$.
   If $|\boldsymbol{\hat{\Sigma}}_1| \neq 0$ define the relative distances $d_1(i) := d(\boldsymbol{x}_i,\boldsymbol{\hat{\mu}}_1,  \boldsymbol{\hat{\Sigma}}_1)$ for $i=1,\dots,n$.
   Now take $H_2$ such that $\{d_1(i) ; i \in H_2\} :=
   \{(d_1)_{1:n},\dots,(d_1)_{h:n}\}$ where $(d_1)_{1:n} \leqslant (d_1)_{2:n}
   \leqslant \dots \leqslant (d_1)_{n:n}$ are the ordered distances, and compute
   $\boldsymbol{\hat{\mu}}_2$
   and $\boldsymbol{\hat{\Sigma}}_2$ based on $H_2$. Then
    $$|\boldsymbol{\hat{\Sigma}}_2| \leqslant
		  |\boldsymbol{\hat{\Sigma}}_1|$$
    with equality if and only if $\boldsymbol{\hat{\mu}}_2 =
       \boldsymbol{\hat{\mu}}_1$ and
   $\boldsymbol{\hat{\Sigma}}_2 = \boldsymbol{\hat{\Sigma}}_1$.
 } % ends italics of theorem.
\vskip0.2in

If $|\boldsymbol{\hat{\Sigma}}_1| > 0$, the C-step thus easily yields a new $h$-subset with lower
covariance determinant. Note that the C stands for `concentration' since $\boldsymbol{\hat{\Sigma}}_2$ is more
concentrated (has a lower determinant) than $\boldsymbol{\hat{\Sigma}}_1$. The condition
$|\boldsymbol{\hat{\Sigma}}_1| \neq 0$ in the theorem is no real restriction because if
$|\boldsymbol{\hat{\Sigma}}_1| = 0$ the minimal objective value is already attained (and in fact the $h$-subset
$H_1$ lies on an affine hyperplane).

C-steps can be iterated until 
$|\boldsymbol{\hat{\Sigma}}_{\text{new}}| =
 |\boldsymbol{\hat{\Sigma}}_{\text{old}}|$. 
The sequence of determinants obtained in this 
way must converge in a finite number of steps 
because there are only finitely many
$h$-subsets, and in practice converges quickly. 
However, there is no guarantee that the final 
value $|\boldsymbol{\hat{\Sigma}}_{\text{new}}|$ 
of the iteration process is the global minimum 
of the MCD objective function. Therefore an 
approximate MCD solution can be obtained by 
taking many initial choices of $H_1$ and 
applying C-steps to each, keeping 
the solution with lowest determinant.

To construct an initial subset $H_1$ one 
draws a random $(p+1)$-subset $J$ and 
computes its empirical mean 
$\boldsymbol{\hat{\mu}}_0$ and covariance 
matrix $\boldsymbol{\hat{\Sigma}}_0$.
(If $|\boldsymbol{\hat{\Sigma}}_0| =0$ then 
$J$ can be extended by adding observations 
until $|\boldsymbol{\hat{\Sigma}}_0| >0$.) 
Then the distances 
$d_0^2(i):= d^2(\boldsymbol{x}_i, 
 \boldsymbol{\hat{\mu}}_0,
 \boldsymbol{\hat{\Sigma}}_0)$ are computed 
for $i=1,\ldots,n$ and sorted. The initial 
subset $H_1$ then consists of the $h$ 
observations with smallest distance $d_0\,$.  
This method yields better initial subsets 
than drawing random $h$-subsets directly, 
because the probability of drawing an 
outlier-free $(p+1)$-subset is much higher 
than that of drawing an outlier-free
$h$-subset.

The FastMCD algorithm contains several computational improvements.
Since each C-step involves the calculation of a covariance matrix,
its determinant and the corresponding distances, using fewer
C-steps considerably improves the speed of the algorithm. It turns
out that after two C-steps, many runs that will lead to the global
minimum already have a rather small determinant.
Therefore, the number of C-steps is reduced by applying only
two C-steps to each initial subset and selecting the 10
subsets with lowest determinants. Only for these 10 subsets further
C-steps are taken until convergence.

This procedure is very fast for small sample 
sizes $n$, but when $n$ grows the computation 
time increases due to the $n$ distances that 
need to be calculated in each C-step. For 
large $n$ FastMCD partitions the data set, 
which avoids doing all calculations on the 
entire data set.

Note that the FastMCD algorithm is itself 
affine equivariant. Implementations of the 
FastMCD algorithm are available in R (as 
part of the packages \textit{rrcov}, 
{\it robust} and {\it robustbase}), in 
SAS/IML Version 7 and SAS Version 9 (in 
\textit{PROC ROBUSTREG}), and in S-PLUS 
(as the built-in function
\textit{cov.mcd}). There is also a Matlab 
version in LIBRA, a LIBrary
for Robust Analysis \cite{Verboven:Toolbox,
Verboven:WIRE-LIBRA} which can be downloaded 
from \url{http://wis.kuleuven.be/stat/robust}\;.
Moreover, it is available in the PLS toolbox 
of Eigenvector Research\linebreak
(\url{http://www.eigenvector.com}). Note that
some MCD functions use $\alpha=0.5$ by default, 
yielding a breakdown value of 50\%, whereas 
other implementations use $\alpha=0.75$. Of 
course $\alpha$ can always be set by the user.

\section*{\sffamily \Large APPLICATIONS} 
There are many applications of the MCD, 
for instance in finance and econometrics 
\cite{Zaman:Econometrics,Welsh:asset,
Gambacciani:mixtures}, 
medicine \cite{Prastawa:brain}, 
quality control \cite{Jensen:controlcharts}, 
geophysics \cite{Neykov:sites},
geochemistry \cite{Filzmoser:expl}, 
image analysis \cite{Vogler:outlier,Lu:brain}
and chemistry \cite{vanHelvoort:geo}, but 
this list is far from complete.

\section*{\sffamily \Large MCD-BASED 
          MULTIVARIATE METHODS} 

Many multivariate statistical methods rely on 
covariance estimation, hence the MCD estimator 
is well-suited for constructing robust 
multivariate techniques. Moreover, the trimming 
idea of the MCD and the C-step have been 
generalized to many new estimators. Here we 
list some applications and extensions.

The MCD analog in regression is the Least 
Trimmed Squares regression estimator 
\cite{Rousseeuw:LMS} which minimizes the sum 
of the $h$ smallest squared residuals
\eqref{eq:LTS}. 
Equivalently, the LTS estimate corresponds to 
the least squares fit of the $h$-subset with 
smallest sum of squared residuals. The FastLTS 
algorithm \cite{Rousseeuw:FASTLTS} uses 
techniques similar to FastMCD. The outlier 
map introduced in \cite{Rousseeuw:Diagnostic} 
plots the robust regression residuals versus 
the robust distances of the predictors, and is 
very useful for classifying outliers, see 
also \cite{Rousseeuw:WIRE-Anomaly}.

Moreover, MCD-based robust distances are also 
useful for robust linear regression 
\cite{Simpson:Onestep,Coakley:Schweppe},
regression with continuous and categorical 
regressors \cite{Hubert:Binary}, and for 
logistic regression 
\cite{Rousseeuw:Sep,Croux:Roblogreg}. In the 
multivariate regression setting (that is, with 
several response variables) the MCD can be used
used directly to obtain 
MCD-regression \cite{Rousseeuw:MCDreg}, whereas 
MCD applied to the residuals leads to 
multivariate LTS estimation \cite{Agullo:MLTS}. 
Robust errors-in-variables regression is 
proposed in \cite{Fekri:orthog}.

Covariance estimation is also important in 
principal component analysis and related 
methods. For low-dimensional data (with 
$n > 5p$) the principal components can be 
obtained as the eigenvectors of the MCD scatter
matrix \cite{Croux:PCAMCD}, and robust factor 
analysis based on the MCD has been studied
in \cite{Pison:RobFA}. 
The MCD was also used for invariant coordinate
selection \cite{Tyler:ICS}.
Robust canonical correlation is proposed 
in \cite{Croux:Cancor}. 
For high-dimensional data, projection pursuit 
ideas combined with the MCD results in the 
ROBPCA method
\cite{Hubert:ROBPCA,Debruyne:IFROBPCA} for 
robust PCA. In turn ROBPCA has led to the 
construction of robust Principal Component 
Regression \cite{Hubert:RPCR} and robust 
Partial Least Squares Regression
\cite{Hubert:RSIMPLS,VandenBranden:PLS}, 
together with appropriate outlier maps, see 
also \cite{Hubert:ReviewHighBreakdown}. Also methods for robust PARAFAC \cite{Engelen:robParafac} and robust multilevel simultaneous component analysis \cite{Ceulemans:RobMSCA} are based on ROBPCA. The LTS subspace
estimator \cite{Maronna:OrReg} generalizes LTS regression to subspace estimation and orthogonal regression. 

An MCD-based alternative to the Hotelling 
test is provided in \cite{Willems:Hotelling}. 
A robust bootstrap for the MCD is proposed 
in \cite{Willems:bootstrapMCD} and a fast 
cross-validation algorithm in 
\cite{Hubert:FastCV}. Computation of the MCD 
for data with missing values is explored in
\cite{Cheng:Missing,Copt:missingsMCD,
Serneels:MissingsPCA}. 
A robust Cronbach alpha is studied in 
\cite{Christmann:Cronbach}. Classification 
(i.e. discriminant analysis) based on MCD is 
constructed in
\cite{Hawkins:Discrim,Hubert:Discrim}, 
whereas an alternative for high-dimensional 
data is developed in
\cite{VandenBranden:RSIMCA}. Robust clustering 
is handled in
\cite{Rocke:Cluster,Hardin:ClusteringMCD,
Gallegos:cluster}.

The trimming procedure of the MCD has inspired 
the construction of maximum trimmed likelihood
estimators
\cite{Vandev:MTL,Hadi:MTL,Muller:Genreg,
Cizek:Binary}, trimmed $k$-means
\cite{Cuesta:kmeans,Cuesta:mixture,
Garcia:linclust}, least weighted squares 
regression \cite{Visek:LWS}, and minimum 
weighted covariance determinant estimation
\cite{Roelant:MWCD}. The idea of the C-step in 
the FastMCD algorithm has also been extended to 
S-estimators \cite{Salibian:FastS,Hubert:DetS}.

\section*{\sffamily \Large RECENT EXTENSIONS}

\section*{\sffamily \large Deterministic MCD}
As the FastMCD algorithm starts by drawing 
random subsets, it does not necessarily give 
the same result at multiple runs of the 
algorithm. (To address this, most 
implementations fix the seed of the random 
selection.)
Moreover, FastMCD needs to draw many initial 
subsets in order to obtain at least one that 
is outlier-free. To circumvent both problems, 
a deterministic algorithm for robust location 
and scatter has been developed, denoted
as DetMCD \cite{Hubert:DetMCD}. 
It uses the same iteration steps as FastMCD 
but does not start from random subsets.
Unlike FastMCD it is permutation invariant, 
i.e.\  the result does not depend on the order 
of the observations in the data set. 
Furthermore DetMCD runs even faster than 
FastMCD, and is less sensitive to point 
contamination. 

DetMCD computes a small number of 
deterministic initial estimates, followed by 
concentration steps. 
Let $X_j$ denote the columns of the data 
matrix $\bX$.
First each variable $X_j$ is standardized by 
subtracting its median and dividing by the $Q_n$ 
scale estimator of \cite{Rousseeuw:Scale}. This 
standardization makes the algorithm location 
and scale equivariant, i.e.\ equations 
\eqref{eq:affine} hold for any non-singular
diagonal matrix $\bA$.
The standardized data set is denoted as the 
$n \times p$ matrix $\bZ$ with rows $\bz'_i$ 
($i=1,\ldots,n$) and columns $Z_j$ 
($j=1,\ldots,p$).

Next, six preliminary estimates $\bS_k$ are 
constructed ($k=1,\ldots,6$) for the scatter
or correlation of $\bZ$:
\begin{enumerate}
\item[1.] $\bS_1=\textrm{corr}(\bY)$ with 
  $Y_j=\textrm{tanh}(Z_j)$ for $j=1,\ldots,p$.  
\item[2.] Let $R_j$ be the ranks of the column 
  $Z_j$ and put $\bS_2= \textrm{corr}(\bR)$. 
	This is the Spearman correlation matrix 
	of $\bZ$.
\item[3.] $\bS_3=\textrm{corr}(\bT)$ with 
  the normal scores 
	$T_j=\Phi^{-1}((R_j - 1/3)/(n+1/3))$.
\item[4.] The fourth scatter estimate is 
  the spatial sign covariance matrix 
	\cite{Visuri:rank}: define 
	$\bk_i=\bz_i/\| \bz_i \|$ for all $i$ 
	and let 
	$\bS_4 =(1/n)\sum_{i=1}^n \bk_i \bk_i'\,$.
\item[5.] $\bS_5$ is the covariance matrix 
  of the $\left\lceil n/2 \right\rceil$ 
	standardized observations $\bz_i$ with 
	smallest norm, which corresponds to the 
	first step of the BACON algorithm 
	\cite{Billor:Bacon}.
\item[6.] The sixth scatter estimate is the 
  raw orthogonalized Gnanadesikan-Kettenring 
	(OGK) estimator \cite{Maronna:OGK}.
\end{enumerate}
As these $\bS_k$ may have very inaccurate 
eigenvalues, the following steps are applied 
to each of them: 
\begin{enumerate}
\item Compute the matrix $\bE$ of eigenvectors 
  of $\bS_k$ and put $\bV = \bZ \bE\,$.
\item Estimate the scatter of $\bZ$ by 
  $\bS_{k}(\bZ) = \bE \bLambda \bE'$ where 
	$\bLambda = \text{diag}(Q^2_n(\bV_1),
	\ldots,Q^2_n(\bV_p))\,$.
\item Estimate the center of $\bZ$ by
  $\hat{\bmu}_k(\bZ) = \bS^{1/2}_k
	(\mbox{comed}(\bZ \bS^{-1/2}_k ))$ where
	$\mbox{comed}$ is the coordinatewise median.
\end{enumerate}
For the six estimates 
$(\hat{\bmu}_k(\bZ),\bS_k(\bZ))$ the 
statistical distances 
$d_{ik} = d(\bz_i,\hat{\bmu}_k(\bZ),
 \bS_k(\bZ))$ of all points are computed as 
in \eqref{eq:d}. 
For each initial estimate $k=1,\ldots,6$ we
compute the mean and covariance matrix of 
the $h_0 = \lfloor n/2\rfloor$ observations 
with smallest $d_{ik}\,$, and relative to
those we compute statistical distances 
(denoted as $d^*_{ik}$) of all $n$ points.
For each $k=1,\ldots,6$ the $h$ observations 
$\bx_i$ with smallest $d^*_{ik}$ are 
selected, and C-steps are applied to them 
until convergence. 
The solution with smallest determinant is 
called the raw DetMCD. Then a weighting step 
is applied as in \eqref{eq:rewMCD}, yielding 
the final DetMCD.

DetMCD has the advantage that estimates can 
be quickly computed for a whole range of 
$h$ values (and hence a whole range of 
breakdown values), as only the C-steps in the 
second part of the algorithm depend on $h$. 
Monitoring some diagnostics (such as the 
condition number of the scatter estimate) can 
give additional insights in the underlying 
data structure, as in the example in 
\cite{Hubert:DetS}.    

Note that even though DetMCD is not affine 
equivariant, it turns out that its deviation 
from affine equivariance is very small. 

\section*{\sffamily \large Minimum 
  regularized covariance determinant}
In high dimensions we need a modification
of MCD, since the existing MCD algorithms 
take long and are less robust in that case. 
For large $p$ we can still make a rough 
estimate of the scatter as follows.
First compute the first $q < p$ robust 
principal components of the data.
For this we can use the MCD-based ROBPCA 
method \cite{Hubert:ROBPCA}, which requires
that the number of components $q$ be set
rather low.
The robust PCA yields a center $\hbmu$ and 
$q$ loading vectors. 
Then form the $p \times q$ matrix $\bL$ 
with the loading vectors as columns. 
The principal component scores $\bt_i$ 
are then given by
   $\bt_i = \bL'(\bx_i - \bmu)\,$.
Now compute $\lambda_j$ for $j=1,\ldots,q$
as a robust variance estimate of the $j$-th
principal component, and gather all the
$\lambda_j$ in a diagonal matrix $\bLambda$. 
Then we can robustly estimate the scatter
matrix of the original data set $\bX$ by 
$\hbSigma(\bX) = \bL \bLambda \bL'$.
Unfortunately, whenever $q < p$ the 
resulting matrix $\hbSigma(\bX)$ will have
$p-q$ eigenvalues equal to zero, hence 
$\hbSigma(\bX)$ is singular.

If we require a nonsingular scatter matrix we
need a different approach using regularization.
The {\it minimum regularized covariance
determinant} (MRCD) method \cite{Boudt:MRCD}
was constructed for this purpose, and works
when $n < p$ too. The MRCD minimizes 
\begin{equation}
  \text{det}\{\rho \bT +
  (1-\rho)\mbox{Cov}(\bX_H)\} 
 \label{eq:MRCD}
\end{equation}
where $\bT$ is a positive definite `target' 
matrix and $\mbox{Cov}(\bX_H)$ is the usual
covariance matrix of an $h$-subset
$\bX_H$ of $\bX$.
Even when $\mbox{Cov}(\bX_H)$ is singular by
itself, the combined matrix is always 
positive definite hence invertible.
The target matrix $\bT$ depends on the
application, and can for instance be the 
$p \times p$ identity matrix or an 
equicorrelation matrix in which the single
bivariate correlation is estimated robustly 
from all the data.
Perhaps surprisingly, it turns out that the
C-step theorem can be extended to the MRCD.
The MRCD algorithm is similar to the 
DetMCD described above, with deterministic
starts followed by iterating these modified
C-steps. The method simulates well even in
1000 dimensions.

Software for DetMCD and MRCD is available from \url{http://wis.kuleuven.be/stat/robust} .

\section*{\sffamily \Large CONCLUSIONS}
In this paper we have reviewed the Minimum 
Covariance Determinant (MCD) estimator of 
multivariate location and scatter.
We have illustrated its resistance to 
outliers on a real data example. 
Its main properties concerning robustness, 
efficiency and equivariance were described, 
as well as computational aspects. We have
provided a detailed reference list with 
applications and generalizations of the MCD 
in applied and methodological research. 
Finally, two recent modifications of the MCD 
make it possible to save computing time and 
to deal with high-dimensional data.

%\bibliographystyle{amsplain}
%\bibliography{Robustness}

% Below was copied from the .bbl file
% (and then expanded):

\end{document}